\documentclass[12pt]{article}

\usepackage{amsfonts}
\usepackage{amsmath}
\usepackage{hyperref}
\usepackage{cite}
\usepackage{epsfig}
\usepackage{latexsym}
\usepackage{paralist}
\usepackage{graphicx}
\numberwithin{equation}{section}
\usepackage[vcentermath]{youngtab}
\usepackage{young}
\usepackage{etex}
\usepackage{braket}
\usepackage{float}
\usepackage{ascmac}

\def\spa#1{\phantom{\fbox{\rule[-#1cm]{0cm}{0cm}}}}

\def\be{\begin{equation}}
\def\ee{\end{equation}}
\def\bea{\begin{eqnarray}}
\def\eea{\end{eqnarray}}

\renewcommand{\thefootnote}{\fnsymbol{footnote}}

\begin{document}

\hfuzz=100pt
\title{{\Large \bf{3d duality with adjoint matter from 4d duality}}}
\author{ 
Keita Nii$^a$\footnote{nii@th.phys.nagoya-u.ac.jp}
}

\maketitle

\vspace*{-1cm}
\begin{center}
$^a${{\it Department of Physics}}
\\ {{\it Nagoya University, Nagoya 464-8602, Japan}}
\spa{0.5}  
\end{center}

\begin{abstract}
We study the Seiberg dualities with an adjoint matter for the $U(N)$ and $SU(N)$ gauge groups in three- and four-dimensions with four supercharges. 
The relation between three- and four-dimensional dualities is investigated.
We especially derive the three-dimensional duality from four-dimensional one by the dimensional reduction including the non-perturbative effect of the $\mathbb{S}^1$-compactification. 
In the $U(N)$ case, we obtain the Kim-Park duality, which is known as a generalization of the Aharony duality including an adjoint matter. In the $SU(N)$ case, we obtain the duality which follows from un-gauging the $U(N)$ Kim-Park duality.    
\end{abstract}

\renewcommand{\thefootnote}{\arabic{footnote}}
\setcounter{footnote}{0}

\newpage

\section{Introduction}\label{Introduction}
Seiberg duality \cite{Seiberg:1994pq} is one of the most interesting properties of the supersymmetric gauge theories since it appears only from the strongly coupled dynamics which is not analyzed perturbatively. Since it is an IR duality, the system which we concern can be described in two ways. One of them sometimes is very useful description for us. 

Though Seiberg dualities are constructed in diverse space-time dimensions, we are especially interested in the three- and four-spacetime dimensions because our world is 4d and 4d physics are sometimes well-understood in terms of 3d physics (for example, see \cite{Aharony:1997bx}). In 4d, the Seiberg duality is first constructed in $\mathcal{N}=1$ supersymmetric gauge theory with only the fundamental matters and it is generalized to the theory with various matter fields (see \cite{Intriligator:1995ax} for example and references therein). Especially the 4d Seiberg duality with an adjoint matter was studied in the presence of a superpotential for the adjoint matter \cite{Kutasov:1995np,Kutasov:1995ve}, which is called Kutasov-Schwimmer duality (Another type of the potentail is considered in \cite{Kapustin:1996nb}). In 3d the Seiberg duality is first studied in \cite{Karch:1997ux,Aharony:1997gp} as well as the dualities for Chern-Simons gauge theories \cite{Giveon:2008zn},\cite{Niarchos:2008jb,Niarchos:2009aa}. The duality in \cite{Aharony:1997gp} is now called Aharony duality and the Chern-Simons dualities are called collectively Giveon-Kutasov dualities and the relation between the Aharony dual and the Giveon-Kutasov dual is now revealed \cite{Intriligator:2013lca}. The ralation between 3d and 4d dualities was previously not clear.

Recently a general procedure to obtain 3d dualities from the 4d Seiberg dualities has been constructed \cite{Aharony:2013dha,Aharony:2013kma}. In \cite{Aharony:2013dha,Aharony:2013kma}, the authors claimed that it is important to consider the theory on $\mathbb{R}^3 \times \mathbb{S}^1$, to take into account the nonperturbative dynamics from the effect of the compactification to $\mathbb{R}^3 \times \mathbb{S}^1$ and to take a low-energy limit with the relations $E \ll \Lambda, \tilde{\Lambda},1/r$ kept in order to derive the corresponding 3d duality, where $\Lambda,\tilde{\Lambda}$ are the dynamical scales in the electric and magnetic sides respectively and $r$ is the radius of the circle. As an example they applied it to the conventional Seiberg duality \cite{Seiberg:1994pq} in which only the fundamental matters are included. They found that the 3d $SU(N_c)$ SQCD is dual to the $U(N_c)$ gauge theory with some fields contents. They also analyzed the dualities for the $Sp(N_c)$ and $SO(N_c)$ gauge groups.

However such a general procedure above is not directly applied to the generic Seiberg duality in 4d since including the various matter fields, for instance an adjoint, symmetric or anti-symmetric matters etc., makes the structure of the Coulomb branch even complicated. 

The 3d dualities with an adjoint matter are studied in \cite{Kim:2013cma,Park:2013wta}. In \cite{Kim:2013cma}, the dual of the 3d $\mathcal{N}=2$ supersymmetric $U(N)$ gauge theory with the superpotential $W=\mathrm{Tr} X^{k+1}$ is constructed and it is called Kim-Park duality. This is derived by generalization of the Aharony duality \cite{Aharony:1997gp}. The Aharony duality can be seen as the 3d counterpart to the 4d conventional Seiberg duality but for $U(N_c)$ gauge group. The 3d $SU(N)$ duality with an adjoint matter is studied in \cite{Park:2013wta}, in which the duality is obtained by using the un-gauging technique \cite{Aharony:2013dha}. Although the dualities with an adjoint matter are known for 3d and 4d, the relation between them is obscure.


In this paper the relation between the 3d and 4d dualities with adjoint matter is investigated especially with focus on the $U(N_c)$ and $SU(N_c)$ gauge groups. We study the Seiberg duality with an adjoint matter on $\mathbb{R}^3 \times \mathbb{S}^1$ carefully, study the non-pertubative effect of the compacttification and derive the corresponding 3d duality by using the 4d Kutasov-Schwimmer duality. In $U(N_c)$ case, we find the dual description of the Kim-Park duality.  In $SU(N_c)$ case we find the duality by using the mirror description \cite{Intriligator:1996ex,deBoer:1997ka,Aharony:1997bx}. The obtained dualities are precisely the same as one obtained by \cite{Kim:2013cma,Park:2013wta}. 

The organization of this paper is as follows. 
In section 2 we review the basic ingredients in 3d $\mathcal{N}=2$ gauge theories and how to derive generically 3d dualities from 4d dualities. 
In section 2 and section 3 we consider the specific case of the 3d dual from 4d dual with an adjoint matter for $U(N)$ and $SU(N)$ gauge groups. 
In section 4 we summarize what we found and discuss the future directions and open problems.

\section{Review of the 3d theories and reduction of the 4d theories}
In this section we will give the basic properties of the $\mathcal{N}=2$ supersymmetric gauge theories in 3d and review how to derive the 3d dualities from 4d dualities. 

We first review the 3d $\mathcal{N}=2$ gauge theories which are obtained by dimensional reduction of the 4d $\mathcal{N}=1$ supersymmetric gauge theories which have four supercharges. The moduli spaces of the 3d $\mathcal{N}=2$ gauge theories are composed of the Higgs branch and Coulomb branch. The Higgs branch is described by the gauge singlet composite operators as mesons and baryons which are constrained both classically and in a quantum way. If we include the adjoint matter, the Coulomb branch which is parametrized by this adjoint squark should be taken into account. The additional Coulomb moduli which are absent in $4d$ $\mathcal{N}=1$ SQCD appear due to the adjoint scalars in a 3d vector superfield. The vector superfields in 3d contain the adjoint scalars due to the gauge fields in 4d along the compactified direction $A_3=\sigma$. The 3d photon is dual to the real compact scalar
\begin{align}
\partial_\mu a \sim \epsilon_{\mu \nu \rho} F^{\nu \rho}.
\end{align}
Thus we describe the additional Coulomb branch as the complex scalar field:
\begin{align}
V=\exp\left( {\frac{2 \pi \sigma }{e^2_3} }+ia \right),
\end{align}
where $e_3$ is the 3d $U(1)$ gauge coupling.
For the $U(N_c)$ gauge group, we classically have the following set of the coordinates:
\begin{align}
V_{i} \sim \exp \left( \frac{\sigma_i}{g_3^2}+i a_i \right), ~~i=1,\cdots,N_c,
\end{align}
while for the $SU(N_c)$ gauge group we have
\begin{align}
V_j \sim \exp \left( \frac{\sigma_j-\sigma_{j+1}}{{g'}_3^{2}}+i( a_j- a_{j+1}) \right), ~~j=1,\cdots,N_c-1,
\end{align}
where $g_3,g'_3$ are the 3d gauge couplings for $U(N_c)$ and $SU(N_c)$ respectively.
Due to the quantum effect, part of these coordinates are lifted, hence the unlifted quantum Coulomb branch is described by
\begin{align}
U(N_c)&:~
\begin{cases}
V_{+} \sim\exp \left( \frac{\sigma_i}{g_3^2} +ia_1   \right)   \\
V_{-} \sim\exp \left( \frac{\sigma_{N_c}}{g_3^2} +ia_{N_c}   \right) 
  \end{cases}
\\  SU(N_c) &:~V=\prod_{j=1}^{N_c-1} V_j \sim \exp \left(  \frac{\sigma_1-\sigma_{N_c}}{g_3^2}+i( a_1- a_{N_C}) \right)
\end{align}
In the presence of the adjoint matter, we have the independent monopole operators \cite{Kim:2013cma}:
\begin{align}
U(N_c): &~~V_{j,\pm} =V_{\pm} \left( X_{11} \right)^{j} \\
SU(N_c): &~~V_{i,j} =V  \left( X_{11} \right)^{i}  \left( X_{N_c N_c} \right)^{j}
\end{align}
where $X_{11}$ and $X_{N_cN_c}$ are the diagonal components of the adoint matter field $X$.
These monopole operators are truncated at $i,j=k-1$ if the superpotential for the adjoint matter $X$ is included:
\begin{align}
W=\mathrm{Tr} \, X^{k+1}.
\end{align}

In contrast to the 4d $\mathcal{N}=1$ SQCD, we can introduce real masses to the chiral superfields aside from the complex masses. This is turned on by gauging the flavor symmetries and adding the vevs for the adjoint scalars of the vector superfields associated with the flavor symmetries. The mass terms for the quarks become
\begin{align}
\bar{\psi}_a^j ((m_{\mathsf{real}})^i_j \delta_{b}^a -\delta_{j}^i \langle \sigma^a_b \rangle ) \psi^b_i,
\end{align}
where $i,j$ are the flavor indices and $a,b$ are the gauge indices.
In the $U(N_c)$ SQCD, since the $U(1)_B$ baryon symmetry is gauged, then we can introduce only the real masses which is traceless. 

In order to connect the 4d dualities with 3d dualities we need to consider the physics on $\mathbb{R}^3 \times \mathbb{S}^1$. Especially the topological solitons on $\mathbb{R}^3 \times \mathbb{S}^1$ do the crucial jobs for the low-energy dynamics connecting the 3d and 4d theories. The instantons on $\mathbb{R}^3 \times \mathbb{S}^1$ are KK-monopoles (or these are called twisted instantons) whose fermionic zero modes are counted by the so-called Callias index theorem \cite{Callias:1977kg,Weinberg:1979zt,Weinberg:1982ev,deBoer:1997kr} and the Atiyah-Singer index theorem \cite{Atiyah:1963zz,Atiyah:1968mp} (Or we can use the $\mathbb{R}^3 \times \mathbb{S}^1$ index theorem \cite{Poppitz:2008hr}):
\begin{align}
N&=N_{\mathsf{instanton}} -N_{\mathsf{monopole}} \\
N_{\mathsf{instanton}}&=2T(r) \\
N_{\mathsf{monopole}}&=\frac{1}{2}\sum_{w \in \mathsf{weights}} \mathrm{sign} \left( w \cdot \sigma \right) w \cdot g
\end{align}
where $\mathrm{Tr} (T^a T^b)=T(r)\delta^{ab}$, $g$ defines the monopole charge and it is a linear combination of the dual simple roots. The summation about $w$ is taken over all the weights in the representation of the field contents. These theorems constrain which types of superpotential are generated dynamically. Namely, if and only if the KK-monopole vertex contains the only two fermionic zero modes, the superpotentials are generated. The superpotentials which appear around the KK-monopole typically contain the Coulomb branch coordinates and some matter fields. In the next section we actually derive the non-perturbative superpotential generated by KK-monopoles in the presence of the adjoint matter.

To obtain the 3d duality we need to take the unusual low-energy limit; $E \ll \Lambda, \tilde{\Lambda}, 1/r $ \cite{Aharony:2013dha}. In such a process we need to include the KK-monopole induced superpotentials. These are crucial steps for obtaining the correct 3d dualities. In the next section we perform explicitly above procedure.

\section{Duality for $U(N_c)$ SQCD with one adjoint matter}
In this section we consider the duality between the $U(N_c)$ SQCD with $N_f$ fundmental flavors and an adjoint matter. We first review the 4d and 3d dualities shrotly and study the non-perturbative superpotential arising from putting these 4d theories on a circle. After that we will derive the 3d duality from the 4d duality.
\subsubsection*{Kutasov-Schwimmer duality}
The 4d duality with an adjoint matter is known as the Kutasov-Shiwimmer duality \cite{Kutasov:1995np,Kutasov:1995ve}. The electric side is a four-dimensional $\mathcal{N}=1$ supersymmetric $SU(N_c)$ gauge theory with $N_f$ fundamental flavors $Q,\tilde{Q}$ and one adjoint matter $X$ with the superpotential $W=\mathrm{Tr} \, X^{k+1}$. The magnetic theory is a four-dimensional $\mathcal{N}=1$ supersymmetric $SU(kN_f -N_c)$ gauge theory with $N_f$ flavors $q,\tilde{q}$, one adjoint matter $Y$ and singlets $M_j \,(j=0,\cdots,k-1)$ with the superpotential $W=\mathrm{Tr} \, Y^{k+1} +\sum_{j=0}^{k-1}M_j \tilde{q} Y^{k-1-j} q$, where $M_j$ are the meson fields identified with the operators on the electric side as $M_j =\tilde{Q} X^j Q$.

We can easily obtain the $U(N_c)$ duality by gauging the $U(1)_B$ baryon flavor symmetry. Since in 4d the dynamics of the $U(1) \subset U(N_c)$ are IR free, then it does not affect the Seiberg duality.

\subsubsection*{Kim-Park duality}
The three-dimensional counterpart of the Kutasov-Schwimmer duality is known as the Kim-Park duality \cite{Kim:2013cma}.
The ``electric" theory is the three-dimensional $\mathcal{N}=2$ supersymmetric $U(N_c)$ gauge theory with $N_f$ fundamental vector-like matters $Q, \widetilde{Q}$ and an adjoint matter $X$ and with the superpotential $W=\mathrm{Tr} \,X^{k+1}$. This theory is simply given by the dimensional reduction of the electric side of the Kutasov-Schwimmer duality.

The ``magnetic" theory is the three-dimensional $\mathcal{N}=2$ supersymmetric $U(k N_f-N_c)$ gauge theory with $N_f$ fundamental vector-like matters $q, \tilde{q}$ and an adjoint matter $Y$ and the singlet $M_j\, (j=0,\cdots,k-1)$ and $v_{0,\pm},\cdots,v_{k-1,\pm}$  with the superpotential $W=\mathrm{Tr}\, Y^{k+1} + \sum_{j=0}^{k-1} M_j  \tilde{q} Y^{k-1-j} q +\sum_{j=0}^{k-1} (v_{j, +} \tilde{v}_{k-1-j,-} +v_{j, -} \tilde{v}_{k-1-j,+}) $, where $v_{0,\pm}$ and $\tilde{v}_{0,\pm}$ are the minimal bare monopole operators of electric side and magnetic side respectively. $v_{j \neq 0,\pm}$ and $\tilde{v}_{j \neq 0,\pm}$ are the monopole operators dressed by the adjoint matter. The Coulomb branches of the electric side are described by the elementary chiral superfields in the magnetic side.

The global charges are summarized as follows, which is applicable in the 4d theory except for the monopole operators. We keep the R-charges generic. Note that $M_j$ are identified with the operator in the electric side as $M_j \sim \tilde{Q} X^{j} Q  $ and we listed the four dimensional instanton factors for a later purpose. These instantons only appear in the 4d theories.
\begin{table}[h]\caption{Quantum numbers of the Kim-Park duality} 
\begin{center}
  \begin{tabular}{|c|c|c|c|c|c| } \hline
     & $SU(N_f)_L$ & $SU(N_f)_R$ & $U(1)_A$ & $U(1)_J$ & $U(1)_R$  \\  \hline 
    $Q$  & $N_f$ & 1 & 1 & 0 & $r$  \\
    $\tilde{Q}$  & 1  & $\bar{N}_f$ & 1 & 0 & $r$   \\
   $X, Y$ & 1 & 1& 0 & 0 & $\frac{2}{k+1}$  \\
   $M_j$ & $N_f$ & $\bar{N}_f$ & 2 &  0 & $2r +\frac{2j}{k+1}$ \\
   $v_{j,\pm}$  & 1  & 1 & $-N_f$ & $\pm 1$ & $-N_f r +N_f -\frac{2}{k+1} (N_c -1) +\frac{2j}{k+1}$  \\ \hline 
   $q$ &  $\bar{N}_f$ &  1  & $-1$ & 0  &  $-r +\frac{2}{k+1}$   \\ 
    $\tilde{q}$ & 1 & $N_f$  & $-1$ & 0 &   $-r +\frac{2}{k+1}$  \\
   $\tilde{v}_{j,\pm}$ & $1$ & $1$ & $N_f$  & $\pm 1$  & $N_fr -N_f + \frac{2}{k+1}(N_c+1) +\frac{2j}{ k+1}   $  \\
   $\eta:= \Lambda^b  $ & 1 & 1 & $2N_f$ & 0 & $2N_f r-2N_f +\frac{4}{k+1} N_c$ \\ 
   $ \tilde{\eta} :=\tilde{\Lambda}^{\tilde{b}}  $ & 1 & 1 & $-2N_f$ & 0 & $-2N_f r+2N_f -\frac{4}{k+1} N_c $   \\ \hline
  \end{tabular}
  \end{center}\label{Kim-Park_electric}
\end{table}
\newline
Here we denote $b,\tilde{b}$ as the one-loop beta function coefficients on electric and magnetic sides respectively and $\Lambda$ and $\tilde{\Lambda}$ are the dynamical scales on both sides.

\subsubsection*{The superpotential induced by the Kaluza-Klein monopole}
If we put the 4d theory on the $\mathbb{S}^1 \times \mathbb{R}^3$, we should include the additional superpotential in the low energy dynamics. 
The symmetry argument and counting the fermionic zero modes imply that the following superpotentials are induced by the ``twisted instanton" which is usually called a Kaluza-Klein (KK) monopole.
\begin{align}
W_{\mathsf{electric}}&= \sum_{\substack{i+j=k-1 \\i,j=0,\cdots,k-1}} \eta \, v_{i,+} v_{j,-} \label{KK_superpotential} \\
W_{\mathsf{magnetic}}&= \sum_{\substack{i+j=k-1 \\i,j=0,\cdots,k-1}} \tilde{\eta} \, \tilde{v}_{i,+} \tilde{v}_{j,-}
\end{align}
The Kaluza-Klein monopoles has the too many fermionic zero modes to appear in the superpotential but they does not have any fundamental quark zero mode. This is a crucial point for deriving the potential. This statement can be easily checked by the Callias index theorem and Atiyah-Singer index theorem (or $\mathbb{R}^3 \times \mathbb{S}^1$ index theorem) \cite{Callias:1977kg,Weinberg:1979zt,Weinberg:1982ev,deBoer:1997kr,Atiyah:1963zz,Atiyah:1968mp,Poppitz:2008hr}.

The index theorems state that the fundamental quark zero mode is absent and only the adjoint fermonic zero modes contribute. Each adjoint field has two fermonic zero modes around the KK-monopole. Since there are two types of adjoint fermion in this theory, the gaugino and the adjoint fermion coming from the adjoint chiral superfield $X$, we have the four fermionic zero modes. These are too many fermonic zero modes to appear in the superpotential. 

However we have the superpotential for the adjoint field $X$ as $W=\mathrm{Tr} \, X^{k+1}$, which generates the potential


%
\begin{align}
V \ni X_{11}^{k-1} \psi_{X,1a}  \psi_{X,a 1},
\end{align}
which is a scalar-fermion interaction. 
This vertex is used to contract the fermionic zero modes arising from the adjoint field $X$. In the end the KK-monopole vertex has only two fermonic zero modes from the gauginos, which can contribute to the superpotential like \eqref{KK_superpotential}.
\begin{align}
\mbox{KK-monopole vertex}:\mathrm{e}^{-S_0} \mathrm{e}^{\sigma +ia} \lambda^2 \psi_X^2 \rightarrow \mathrm{e}^{-S_0} \mathrm{e}^{\sigma +ia} \lambda^2 \left(X_{11} \right)^{k-1}
\end{align}
%

\subsubsection*{The weak perturbation}
In deriving the 3d duality, it is helpful to consider the weak perturbation of the both theory by the potential for the adjoint matter $X,Y$, which breaks the $U(1)_R$ symmetry explicitly generating a new $U(1)_R$ symmetry at low energy.
We consider the following deformation of the superpotential \cite{Kutasov:1995np}:
\begin{align}
W=\sum_{j=0}^{k} \frac{s_j}{k+1-j} \mathrm{Tr}\, X^{k+1-j} \label{perturbation}
\end{align}
The minima of the superpotential are following.
\begin{align}
W' (x) = \sum_{j=0}^{k} s_j x^{k-j} =s_0 \prod_{j=1}^{k} (x-a_j)
\end{align}
When the $\left\{ a_j\right\}$ are distinct, all the components of the adjoint field become massive (This is easily checked by shifting the field as $X \rightarrow \langle X \rangle +\delta X$). The gauge symmetry is broken as
\begin{align}
U(N_c) \rightarrow U(i_1) \times \cdots \times U(i_k),~~\sum_{a=1}^{k} i_a=N_c.
\end{align}

The low energy effective theory is the sum of the $\mathcal{N}=2$ $U(i_a)$ theory with $N_f$ fundamental quarks with no  adjoint matter and no superpotential. Thus we can use the same argument as the derivation of the duality of 3d SQCD with no adjoint matter. The monopole operators in the electric side at high energy $v_{j,\pm}~(j=0,\cdots,k-1)$ are correctly describing the Coulomb branches of the $U(i_1) \times \cdots \times U(i_k)$ gauge groups at low energy. Then we can identify them as
\begin{align}
v_{j,\pm} &\sim \mbox{linear~combination of~} v^{low~energy}_{U(i_{j})} \nonumber \\
 v^{low~energy}_{U(i_{j})}&:~ \mbox{Coulomb~branch~coordinates~of~the~}U(i_{j})  \nonumber
\end{align}
This weak deformation helps us to study the duality especially on the magnetic side as well as the $SU(N_c)$ duality in Section 4.

\subsection*{The Kim-Park dual from 4d}
We will here derive the Kim-Park duality using the 4d Kutasov-Schwimmer duality with $U(N_c)$ gauge group. We straightforwardly construct the duality on $\mathbb{R}^3 \times \mathbb{S}^1$ by adding the superpotential generated by the KK-monopoles to the 4d duality. Next we do the mass deformation to the theory in order to obtain the duality without $\eta$ terms. We show how to do it in both electric and magnetic theories respectively.




\subsubsection*{The electric theory}  
First we put the electric theory on a circle.
The theory is $U(N_c)$ gauge theory with the $N_f$ fundamental matters $Q,\,\widetilde{Q}$, one adjoint field $X$ with the superpotential $W=\mathrm{Tr} \, X^{k+1} +\sum_{i+j=k-1} \eta \, v_{i,+} v_{j,-}$, which contain the non-perturbative sperpotential from the effect of $\mathbb{S}^1$ compactification (KK-monopoles). We would like to obtain the theory without $\eta$ term. Then we will start with $N_f+2$ flavors and turn on the real masses for the $SU(N_f+2) \times SU(N_f+2)$ flavor symmetries.
\begin{align}
m=\left(
    \begin{array}{ccccc}
      0 &  &  & & \\
         & \ddots &  & & \\
         &  & 0 & & \\
         &  &     & m &  \\
         & & &  & -m
    \end{array}
  \right), ~~
  \tilde{m}=\left(
    \begin{array}{ccccc}
      0 &  &  & & \\
         & \ddots &  & & \\
         &  & 0 & & \\
         &  &     & -m &  \\
         & & &  & m
    \end{array}
  \right) \label{real_mass1}
\end{align}
Notice that this is traceless as it should be. Integrating out the massive flavors, we obtain at the low energy the three-dimensional $\mathcal{N}=2$ $U(N_c)$ gauge theory with $N_f$ flavors and one adjoint matter with no introduction of the Chern-Simons terms. The global charges are precisely those of the Kim-Park duality at Table \ref{Kim-Park_electric}. The monopole operators in \eqref{KK_superpotential} vanish at low energy because they are pinched due to the low-energy $U(1)$ dynamics \cite{Aharony:1997bx}. 

If we turn on the potential \eqref{perturbation}, then we obtain the sum of the $U(i_j)$ gauge theory with $N_f+2$ flavors and no adjoint matter at low energy. In the presence of the real mass deformations \eqref{real_mass1} we can take the low energy limit at $Q^{U(i_j)}_{N_f+1}=Q^{U(i_j)}_{N_f+2}=\tilde{Q}^{U(i_j)}_{N_f+1}=\tilde{Q}^{U(i_j)}_{N_f+2}=0$ and obtain the sum of the $U(i_j)$ gauge theory with $N_f$ flavors and no adjoint matter, where $Q^{U(i_j)}$s mean the $U(i_j)$ fundamental quarks. In the limit of $s_j =0 ~(j \neq 0)$, we expect that the $U(N_c)$ gauge theory with $N_f$ flavors and one adjoint field with the superpotential $W=\mathrm{Tr} \, X^{k+1}$ recover.

\subsubsection*{The magnetic theory with $\eta$ term}
Next we will put the dual theory on a circle and do the same deformation as the electric side.
On the dual side we have the $U(k (N_f+2)-N_c)$ gauge theory with the $N_f+2$ fundamental matters $q,\,\tilde{q}$, one adjoint field $Y$ and the singlets $M_j\,(j=0,\cdots,k-1)$ with the superpotential $W=\mathrm{Tr}\, Y^{k+1} + \sum_{j=0}^{k-1} M_j \tilde{q} Y^{k-1-j} q +\sum_{j=0}^{k-1} \tilde{\eta} \, \tilde{v}_{j,+} \tilde{v}_{k-1-j,-}$.

Since the real mass deformation in the electric side \eqref{real_mass1} is the $SU(N_f+2)\times SU(N_f+2)$ background gauging, it is easily mapped to the dual side. The real masses in the magnetic theory are 
\begin{align}
m_{\mathsf{dual}} =\left(
    \begin{array}{ccccc}
      0 &  &  & & \\
         & \ddots &  & & \\
         &  & 0 & & \\
         &  &     & -m &  \\
         & & &  & m
    \end{array}
  \right), ~~ \tilde{m}_{\mathsf{dual}}=\left(
    \begin{array}{ccccc}
      0 &  &  & & \\
         & \ddots &  & & \\
         &  & 0 & & \\
         &  &     & m &  \\
         & & &  & -m
    \end{array}
  \right).
\end{align}
We need to find the correct vacuum which corresponds to the vacuum of the electric side and take a low-energy limit. It helps us to use the weak deformation \eqref{perturbation}. 
With the introduction of the weak deformation, the dual gauge group is broken in two steps. At first the gauge group is broken by the deformation of the theory by $W=\sum_{j=0}^{k} \frac{s_j}{k+1-j} \mathrm{Tr} \, Y^{k+1-j} $;
\begin{align}
U(k(N_f+2) -N_c) \rightarrow U(r_1) \times \cdots \times U(r_k),~~\sum_{i=1}^{k} r_i =kN_f-N_c +2k \label{dual_gauge_breaking}
\end{align}
In this breaking all the components of the adjoint matter become massive and integrated out.
At the second step, the gauge group is broken by the vevs of the $\tilde{A}_3=\tilde{\sigma}$ adjoint scalar field in the vector superfield.
\begin{align}
&\widetilde{\sigma}_{U(r_i)\,\mathsf{part}} =  \left(
    \begin{array}{ccccc}
         0 &    &   &  &  \\
         & \ddots &   &  &  \\
       &  &  0 & &  \\
        & & & -m &  \\
        & & &      & m  
     \end{array}
  \right) 
  \end{align}
  \begin{align}
U(k(N_f+2) -N_c)&\rightarrow U(r_1) \times \cdots \times U(r_k),~~~~~~\sum_{i=1}^{k} r_i =kN_f-N_c +2k \nonumber \\
& \rightarrow (U(r_1 -2) \times U(1)^2 )\times  \cdots \times ( U(r_k-2)\times U(1)^2 )
\end{align}
At the low energy the adjoint chiral superfield $Y$ is massive and the low energy effective theory is the sum of the $U(r_i-2) \times U(1)^2$ theory with fundamental matter with meson singlets. The fundamental matters and the mesonic fields decompose as follows.

\begin{itemize}
\item $\hat{q}^{U(r_i -2)}_f,\hat{\tilde{q}}^{U(r_i -2)}_{\tilde{f}}$~$(f,\tilde{f}=1,\cdots,N_f;~~i=1,\cdots,k)$  :\\$U(r_i-2)$ fundamental quarks which are $N_f$ flavors. The $f,\tilde{f}$ are flavor indices.
\item $q^{a}_i,\tilde{q}^{a}_i$~$(a=1,2~~i=1,\cdots,k) $:\\ $\mathbf{1}_{\pm\delta_{a1},\pm\delta_{a2}}$ representation under the gauge group $U(r_i-2) \times U(1)^2$. The lower indices indicate the corresponding $U(1)$ charges. 
\item $\hat{M}_j~(j=0,\cdots,k-1)$ : \\
 $N_f \times N_f$ mesons which come from the left-upper components of the meson chiral superfields $M_j$. 
\item $M_{aj} ~(a=1,2;~j=0,\cdots,k-1)$ :\\
 Singlets coming from the $N_f$-th and $(N_f+1)$-th components of the mesons $M_j$. $M_{1j} \equiv M_{N_f+1,N_f+1},~M_{2j} \equiv M_{N_f+2,N_f+2}$
\end{itemize} 
We define the composite operators;
\begin{align}
\hat{N}_{i}^j & :=\hat{\tilde{q}}^{U(r_i -2)} Y^{j}  \hat{q}^{U(r_i -2)} \\
N_{ai}^j &:= \tilde{q}^{a}_i Y^{j}  q^{a}_i
\end{align}
Strictly speaking, since we are in the broken phase \eqref{dual_gauge_breaking} where we have no adjoint matter $Y$, we should write $Y$ field as the vacuum expectation value $\braket{Y}$. However since we finally take the limit where the weak deformations are switched off and we expect the adjoint fields to recover as the massless degree of freedom, we write the adjoint matter $Y$ naively as the dynamical field. In the following since we deal with $Y$ as the vevs, the upper indixes of 
$\hat{N}_{i}^j ,N_{ai}^j $ does not distinguish the independent chiral operators. The upper indises are only important in turning off the weak deformations.

In addition, we define the monopole operators which describe the Coulomb branches of $ (U(r_1 -2) \times U(1)^2 )\times  \cdots \times ( U(r_k-2)\times U(1)^2 )$:
\begin{align}
\hat{\tilde{V}}_{j,\pm} &:~\mbox{Coulomb branch of the}~U(r_{j+1}-2).~(j=0,\cdots,k-1) \\
\tilde{v}_{U(1)}^{i,a} &:~\mbox{Coulomb branch of the}~U(1)^2.~ (i=1,\cdots,k;~a=1,2)
\end{align}

In this notation we have the superpotential at low energy,

\begin{align}
W &= \sum_{i=1}^k \sum_{j=0}^{k-1} \hat{M}_j \hat{N}_{i}^{k-1-j} +  \sum_{i=1}^k \sum_{j=0}^{k-1}  M_{1j} N_{1,i}^{k-1-j}  +\sum_{i=1}^k \sum_{j=0}^{k-1}  M_{2j}N_{2,i}^{k-1-j}  
        + \sum_{j=0}^{k-1}  \tilde{\eta} \, \tilde{v}_{j,+} \tilde{v}_{k-1-j,-}    \nonumber \\
        &\qquad +\sum_{\substack{ a=1,2 \\ i=1,\cdots,k}  } N_{a,i }^{0} \tilde{v}_{U(1),+}^{i,a} \tilde{v}_{U(1),-}^{i,a}   +\sum_{j=1}^k  \hat{\tilde{V}}_{j,+} \tilde{v}_{U(1),-}^{j,1} + \sum_{j=1}^k \hat{\tilde{V}}_{j,-} \tilde{v}_{U(1),+}^{j,2}  
\end{align}
where the fifth term is due to the $U(1)^2$ dynamics. These $U(1)$ parts are the sum of the $\mathcal{N}=2$ supersymmetric QED with one flavor. Then we can use the dual description of the $\mathcal{N}=2$ XYZ model with the above superpotential \cite{Aharony:1997bx}. The sixth and seventh terms come from the effect of the Affleck-Harvey-Witten type superpotential \cite{Affleck:1982as} through the breaking of $U(r_i) \rightarrow U(r_1-2) \times U(1) \times U(1)  $. $\hat{\tilde{V}}_{j,\pm}$ should be identified with the monopole operators of the $U(kN_f-N_c)$ in the limit $s_j \,(j \neq 0) \rightarrow 0$ and the monopole operators $\tilde{v}_{j,\pm}$ in the high energy should be identified with the linear combinations of the monopole operators $\tilde{v}_{U(1),+}^{j,1},\,\tilde{v}_{U(1),-}^{j,2}$ of the $U(1)^2$ part at low energy.   

The equations of motion drop the second, third, fourth, and fifth terms off. Since the $U(r_i-2)$ gauge groups are combined to $U(kN_f-N_c)$ groups with an adjoint matter $Y$ in the limit of $s_1,\cdots,s_k=0$, we finish with the Kim-Park magnetic dual. The magnetic theory in the end becomes the $U(kN_f-N_c)$ gauge theory with $N_f$ flavors, one adjoint matter and singlets $M_j,~v_{i,\pm}$ with the superpotential
\begin{align}
W=\mathrm{Tr} \, Y^{k+1} + \sum_{j=0}^{k-1} M_j  \tilde{q} Y^{k-1-j} q +\sum_{j=0}^{k-1} (v_{j, +} \tilde{v}_{k-1-j,-} +v_{j, -} \tilde{v}_{k-1-j,+}),
\end{align}
where the $U(1)^2$ Coulomb branch coordinates $ \tilde{v}_{U(1),-}^{j,1},~\tilde{v}_{U(1),+}^{j,2} $ are identified with the chiral superfields $v_{j,\pm}$ in the Kim-Park magnetic theory and the Coulomb branch coordinates of the $U(kN_f-N_c)$ are denoted as $\tilde{v}_{j,\pm}$. Notice that the $U(1)^2$ part was dualized to the theory which only contains the chiral superfields with no gauge symmetry. In the limit of $s_j \,(j \neq 0) \rightarrow 0$, the part with $U(1)^2 \times \cdots \times U(1)^2$ gauge symetry becomes the $U(k)\times U(k)$ gauge theory. Each $U(k)$ theory contains one fundamental matter and an adjoint field. The vacuum of the $N_f=1$ theory usually has a runaway behavior \cite{Aharony:1997bx}. In this case, however, the various terms in the superpotential may stabilize the vacuum. In deriving the above duality we assume that the enhancement of the gauge symmetry to $U(k)$ does not change the duality and the $U(1)$ physics correctly produces the duality. 

\section{Duality for $SU(N_c)$ SQCD with one adjoint matter}
In this section we will derive the three-dimensional duality for $SU(N_c)$ with one adjoint matter from reduction of the 4d duality. We will eventually obtain the 3d duality studied by \cite{Park:2013wta} whose authors used the un-gauging techique in deriving it. The 4d Kutasov-Schwimmer duality which we employ is the same as the previous section but we do not gauge the $U(1)_B$ flavor symmetry. 

\subsection*{Park-Park dual from 4d}
Using the 4d Kutasov-Schwimmer duality and compactifying the both theories on $\mathbb{S}^1$, we obtain the 3d duality. In such a process we need to include the non-perturbative superpotential which arises from the twisted instantons. We also need to deform both theories by the real masses to obtain the theories with no $\eta$ term.

We show the quantum numbers of the field contents in 4d theories and $\mathbb{R}^3 \times \mathbb{S}^1$ theories in Table \ref{2},3. Since in the three-dimensional limit  we have no axial anomaly, we show the additional $U(1)_A$ symmetry in the following table.
The field contents are the same as the four dimensional ones, but in three-dimensions there are Coulomb branches which should be naturally identified with the chiral superfields. We will use the symbol $V,\tilde{V}$ for the Coulomb branch coordinates.

\begin{table}[H] \caption{Quntum numbers of the electric side}  \begin{center}
  \begin{tabular}{|c|c|c|c|c|c|c| } \hline
     & $SU(N_c)$ & $SU(N_f)_L$ & $SU(N_f)_R$ & $U(1)_B$ & $U(1)_A$ & $U(1)_R$ \\  \hline 
    $Q$  & $N_c$ & $N_f$ & 1 & 1 & 1& 0 \\
    $\tilde{Q}$  & $\bar{N_c}$  & 1 & $\bar{N_f}$ & $-1$& 1 & 0  \\
   $X$ & $\mathsf{adj.}$ & 1&1&0&0& $\frac{2}{k+1}$ \\
   $M_j$ &1& $N_f$ & $\bar{N_f}$ &  0 & 2 & $\frac{2j}{k+1}$ \\
   $V_{ij}$ &  1 &  1  & 1 & 0  &  $-2N_f$ & $2 N_f - \frac{4}{k+1} (N_c-1) +\frac{2(i +j)}{k+1}$   \\
   $ \Lambda^b  $ & 1 & 1 & 1 & $0$ & $2N_f$  & $-2N_f +\frac{4N_c}{k+1}$ \\ \hline
  \end{tabular} \end{center} \label{2}
\end{table}
\begin{table}[H]
\begin{center}\caption{Quantum numbers of the magnetic side} \small
  \begin{tabular}{|c|c|c|c|c|c|c| } \hline
     & $SU(kN_f -N_c)$ & $SU(N_f)_L$ & $SU(N_f)_R$ & $U(1)_B$ & $U(1)_A$ & $U(1)_R$ \\  \hline 
    $q$  & ${\tiny\yng(1)}$ & $\bar{N_f}$ & 1 & $\frac{N_c}{k N_f -N_c}$ & $-1$ & $\frac{2}{k+1}$ \\
    $\tilde{q}$  & $ \overline{{\tiny\yng(1)}}$  & 1 & $N_f$ & $-\frac{N_c}{kN_f -N_c}$& $-1$ & $\frac{2}{k+1}$  \\
   $Y$ & $\mathsf{adj.}$ & 1&1&0&0& $\frac{2}{k+1}$ \\
   $M_j$ &1& $N_f$ & $\bar{N_f}$ &  0 & 2 & $\frac{2j}{k+1}$ \\
   $\tilde{V}_{ij}$ &1&1&1&0& $2N_f$ & $ -2 N_f + \frac{4}{k+1} (N_c+1) +\frac{2(i+j)}{k+1} $ \\ 
   $\tilde{\Lambda}^{\tilde{b}}$ &  1 & 1 & 1 & 0 & $- 2N_f$ & $2N_f -\frac{4N_c}{k+1}$ \\ \hline
  \end{tabular}\end{center} \label{8}
\end{table}

\subsubsection*{The monopole operators}
We consider the $SU(N_c)$ gauge theories and the monopole background. We restrict ourself to the part of Coulomb branch which is labeled by $\phi_1 > \phi_2 >\cdots > \phi_K >0 >\phi_{K+1} >\cdots >\phi_{N_c}$, where $\phi_i$ is the adjoint scalar fields of the vector superfields. In this region Callias index theorem tells us that the numbers of the fermonic zero modes around the monopole-instanton are
\begin{align}
&\mbox{quark zero modes: } N_{\Box} =\begin{cases}
    1 & (\phi_K >0 >\phi_{K+1}) \\
    0 & (\phi_i > \phi_{i+1} >0 ~\mbox{or} ~ 0>\phi_i > \phi_{i+1}) 
  \end{cases}  \\
&\mbox{adj. fermion zero modes: } N_{\mathsf{adj.}} =2.
\end{align}
Then the global charges of the monopole operators are as follows,
\begin{table}[H] \caption{Quntum numbers of the monopole operators}
\begin{center}
  \begin{tabular}{|c|c|c|} \hline
     & $U(1)_A$ & $U(1)_R$  \\ \hline 
    $ Y_1 $ & 0 & $-2-2(\frac{2}{k+1} -1) =-\frac{4}{k+1}$ \\
    \vdots & \vdots&  \vdots \\ 
    $Y_{K-1}$ & 0 &  $ -\frac{4}{k+1}$ \\ 
   $Y_K$ & $-2 N_f$ & $ -2-2(\frac{2}{k+1} -1) -2N_f (-1)  =2N_f -\frac{4}{k+1}$ \\
   \vdots & \vdots & \vdots \\ 
   $Y_{N_c-1}$ &  0 & $-\frac{4}{k+1} $  \\
  $ V \equiv \prod_{i=1}^{N_c-1} Y_i $ &  $-2N_f$  &$ 2N_f -\frac{4}{k+1} (N_c -1) $ \\ \hline
\end{tabular}
\end{center}
\end{table}
\noindent where we use the following notations for the bare and dressed monopole operators:
\begin{align}
Y_i &\sim \exp \left( \frac{\Phi_i -\Phi_{i+1}}{g^2} \right) \label{monopoleop} \\
V_{ij} & \sim (X_{11})^i (X_{N_c N_c})^j  V. \label{dressedmonopole}
\end{align}
The chiral superfield $\Phi_i$ consists of the adjoint scalar $\phi_i$ and the dual photon $a_i$, which is defined as $\Phi_i =\frac{\phi_i}{g_{3}^2 } +ia_i$. The tilde in the above monopole operators means that the r.h.s of \eqref{monopoleop} is only legitimate at the semi-classical domain with large $\phi_i$. At the small $\phi_i$, the 3d gauge coupling has a non-trivial loop correction and the metric on the Coulomb branch becomes intricate, so the definition of the monopole operator is involved.
The powers of $X_{11}$ and $X_{N_c N_c}$ in \eqref{dressedmonopole} are truncated at $O(X^{k-1})$ due to the superpotential $W=\mathrm{Tr}\,X^{k+1}$. The relation of the monopole operators between the $SU(N_c)$ and $U(N_c)$ gauge theories with an adjoint matter is simple:
\begin{align}
V &= V_+ V_- \\
V_{ij} &= V_{+i} V_{-j}
\end{align}

\subsubsection*{KK-monopole induced superpotential}
By the symmetry argument we assume the following superpotentials are generated by the 1 KK-monopole (twisted instanton) configuration.
\begin{align}
W_{\mathsf{ele}} &= \sum_{j =0 }^{k-1} \Lambda^b V_{jk-1} = \sum_{j =0 }^{k-1} \eta V_{jk-1-j} \label{SU(N)KK} \\
W_{\mathsf{mag}} &=  \sum_{j =0 }^{k-1}\tilde{\Lambda}^{\tilde{b}} \tilde{V}_{jk-1-j} = \sum_{j =0 }^{k-1} \tilde{\eta} \tilde{V}_{jk-1-j}
\end{align}
The index theorem says that the fermionic zero modes around the KK monopole solution are too many for the KK monopole to contribute to the superpotential. However we now have the adjoint matter which interact via the superpotential as
\begin{align}
W=\mathrm{Tr} \, X^{k+1}.
\end{align}
This superpotential produces the following interaction.
\begin{align}
\left( X_{11} \right)^{j} \psi_{X,1N_c} \left( X_{N_cN_c} \right)^{k-1-j} \psi_{X,N_c1}
\end{align}
Then the KK monopole vertex
\begin{align}
\mathrm{e}^{-S_0} \mathrm{e}^{\phi +i \sigma} \lambda ^2 \psi_{X}^2 
\end{align}
is modified to
\begin{align}
\mathrm{e}^{-S_0} \mathrm{e}^{\phi +i \sigma}  \lambda^2 \left(  X_{11} \right)^j  \left( X_{N_cN_c}\right)^{k-1-j}
\end{align}
which is precisely coming from the above non-perturbative superpotential \eqref{SU(N)KK}.

\subsubsection*{3d Kutasov-Schwimmer duality with no $\eta$ terms.}
As in the $U(N_c)$ duality, we first construct the duality on $\mathbb{R}^3 \times \mathbb{S}^1$. This can be easily done by including the KK-monopole induced superpotential. Next we deform the theory by the real masses in order to obtain the duality without $\eta$ terms.

The electric side is the three-dimensional $\mathcal{N}=2$ supersymmetric $SU(N_c)$ gauge theory with $N_f$ fundamental flavors $Qs,\tilde{Q}s$ and one adjoint field $X$ with the superpotential $W= \mathrm{Tr} \, X^{k+1} +\sum_{i+j=k-1} \eta V_{ij}$. The magnetic side is the three-dimensional $\mathcal{N}=2$ supersymmetric $SU(kN_f -N_c)$ gauge theory with $N_f$ fundamental flavors $qs,\tilde{q}s$ and one adjoint field $Y$ with the superpotential $W=\mathrm{Tr} \, Y^{k+1} +\sum_{j=0}^{k-1} M_j \tilde{q} Y^{k-1-j} q +\sum_{i+j=k-1} \tilde{\eta} \tilde{V}_{ij}$. 

In order to obtain the duality without the $\eta$ terms we start with the $N_f +1$ flavors as the $U(N_c)$ case. We turn on the real mass ``$m$'' for the last flavor in the electric theory side, which is a background gauging of the $SU(N_f+1) \times SU(N_f+1) \times U(1)_B$ flavor symmetries. In the dual side the corresponding real mass is mapped as follows.
\begin{align}
&m_r^{\mathsf{dual}} = \left(
    \begin{array}{cccc}
      m_1 &  &  &  \\
       & \ddots &  &  \\
      &  & m_1 &  \\
       &  &  & m_2
    \end{array}
  \right)  ,~~\widetilde{m}_{r}^{\mathsf{dual}} =\left(
    \begin{array}{cccc}
     - m_1 &  &  &  \\
       & \ddots &  &  \\
      &  & -m_1 &  \\
       &  &  & -m_2
    \end{array}
  \right)  \\
  &m_1=\frac{k}{kN_f +k -N_c}m ,~~
  m_2 =\frac{N_c -k N_f}{k N_f +k -N_c}m
\end{align}

The electric theory flows to the $SU(N_c)$ gauge theory with $N_f$ flavors with no $\eta$ terms since the high energy monopole operators vanish due to the absence of the complex masses as discussed in \cite{Aharony:2013dha}. The quantum numbers of the electric theory are summarized in Table \ref{4},
\begin{table}[H] \begin{center}\caption{Quantum numbers of the 3d $SU(N_c)$ theory on the electric side} \label{4}
  \begin{tabular}{|c|c|c|c|c|c|c| } \hline
     & $SU(N_c)$ & $SU(N_f)_L$ & $SU(N_f)_R$ & $U(1)_B$ & $U(1)_A$ & $U(1)_R$ \\  \hline 
    $Q$  & $N_c$ & $N_f$ & 1 & 1 & 1& $r$ \\
    $\tilde{Q}$  & $\bar{N_c}$  & 1 & $\bar{N_f}$ & $-1$& 1 & $r$  \\
   $X$ & $\mathsf{adj.}$ & 1&1&0&0& $\frac{2}{k+1}$ \\  \hline
   $M_j$ &1& $N_f$ & $\bar{N_f}$ &  0 & 2 & $2r+\frac{2j}{k+1}$ \\
    $V_{ij}$ &  1 &  1  & 1 & 0  &  $-2N_f$ & $2 N_f(1-r)  - \frac{4}{k+1} (N_c-1) +\frac{2(i +j)}{k+1}$   \\ \hline
  \end{tabular}\end{center}
\end{table} 
\noindent where we keep the R-charges of the quarks as generic values.

The flow in the magnetic side is complicated and it is helpful to consider the weak deformation of the theory.
We consider the weak perturbation of the theories by
\begin{align}
W_{\mathsf{ele}}=  \sum_{i=0}^{k} g_{i} \mathrm{Tr}\, X^{i+1} \\
W_{\mathsf{mag}}=  \sum_{i=0}^{k} g_{i} \mathrm{Tr}\, Y^{i+1} 
\end{align}
where $g_0$ is the Lagrange multiplier imposing the constraint such that $\mathrm{Tr}\, X=0 $. This weakly perturbed theory has generically the k vacua for the vevs of $X$. Then the gauge symmetry is generically broken by the vevs of the adjoint field $X$ like $SU(N_c) \rightarrow SU(i_1) \times SU(i_2) \times \cdots \times SU(i_k)\times U(1)^{k-1}$, where $\sum_{j=1}^{k} i_j =N_c$. 

In the electric side, we start with $N_f +1$ flavors, take $\phi_i \equiv 2 \pi \sigma_i =0$ and flow to the low energy limit with $Q^{N_f+1}$ and $\tilde{Q}^{N_f+1}$ integrated out. By tuning $g_i \rightarrow 0 \,(i \neq k)$ we finally get the $SU(N_c)$ gauge theory with $N_f$ flavors and an adjoint field recovered at the low energy. This is a three-dimensional version of the electric theory in the Kutasov-Schwimmer duality.

In the magnetic side, $W=\sum_{i=0}^{k} g_i \mathrm{Tr}\, Y^{i+1}$ breaks the gauge group as $SU(kN_f-N_c) \rightarrow SU(n_1) \times \cdots \times SU(n_k) \times U(1)^{k-1},  \, \sum_{i=1}^k n_i =kN_f-N_c$.  
 When we start with $N_f+1$ flavors and turn on the real masses corresponding to the electric side, we have the additional breaking of the gauge symmetry by the vevs of $\tilde{A_3} \equiv\tilde{\sigma}$ as follows.
\begin{align}
SU(k (N_f+1) -N_c) &\rightarrow SU(N_1 ) \times \cdots \times SU(N_k) \times U(1)^{k-1}   \nonumber  \\ 
&\rightarrow SU(n_1) \times \cdots \times SU(n_k) \times SU(p_1)\times \cdots \times SU(p_k) \nonumber \\
& \qquad \qquad\times U(1)^{k} \times U(1)^{k-1}
\end{align}
where 
\begin{gather}
N_i:=n_i+p_i \\
\sum_{i=i}^k N_i =kN_f-N_c+k ,~~\sum_{i=1}^k n_i =kN_f-N_c,~~\sum_{i=1}^{k} p_i =k 
\end{gather}
\begin{align}
\tilde{\sigma}_{N_1 \times N_1} =\left(
    \begin{array}{rrr}
     a_{11}  &  &  \\
        & \ddots  &  \\
       &  & a_{1N_1}
    \end{array} 
    \right) , \cdots, 
    \tilde{\sigma}_{N_k \times N_k} =\left(
    \begin{array}{rrr}
     a_{k1}  &  &  \\
        & \ddots  &  \\
       &  & a_{kN_k}
    \end{array} 
    \right)
\end{align}
where $a_{ij}$ take the value of $-m_1$ or $-m_2$ and satisfy the condition that $\sum_{i,j} a_{ij} =0$ and $\mbox{the~number~of~}-m_1$ is $kN_f -N_c$ due to the traceless condition of the adjoint field $\tilde{\sigma}$.

We need to choose a vacuum which corresponds to the vacuum of the electric side, which is 
\begin{align}
\tilde{\sigma}_{N_1 \times N_1} =\left(
    \begin{array}{rrrr}
     -m_1  &  &  &\\
        & \ddots  & &  \\
       &  & -m_1 &  \\
       &&& -m_2
    \end{array} 
    \right) , \cdots, 
    \tilde{\sigma}_{N_k \times N_k} =\left(
    \begin{array}{rrrr}
     -m_1  &  & &  \\
        & \ddots   &&  \\
       &  & -m_1 &\\
       &&& -m_2
    \end{array} 
    \right). \label{vacuum}
\end{align}
Then the gauge group is broken as
\begin{align}
SU(k(N_f+1) -N_c)  &\rightarrow SU(N_1-1) \times \cdots \times SU(N_k-1) \times U(1)^{k-1} \times U(1) \times U(1)^{k-1}.
\end{align}

Here we will clarify the $U(1)^{2k-1}$ part in detail. The first $U(1)^{k-1}$ corresponds to the generators coming from the left-upper $(kN_f -N_c ) \times (kN_f -N_c)$ part. The generator of the second $U(1)$ is as follows. 
\begin{align}
T=\left(
    \begin{array}{ccccc}
      1&  &  && \\
       & 1 &  &&\\
       &  & 1 && \\
       &&& -\frac{3}{2}&\\
       &&&& -\frac{3}{2}
    \end{array}
  \right) , ~~\mathrm{Tr} \, T =0
\end{align}
where we show the case with $kN_f-N_c=3,~k=2$ for simplicity. The last $U(1)^{k-1}$ is the right-lower $k \times k$ part:
\begin{align}
T'=\left(
    \begin{array}{ccccc}
      0&  &  && \\
       & 0 &  &&\\
       &  & 0 && \\
       &&& 1&\\
       &&&& -1
    \end{array}
  \right),
\end{align}
which is also the case with $kN_f-N_c=3,~k=2$.

In order to obtain the original $SU(N_c)$ duality with an adjoint matter and with no $\eta$ term, we need to tune the potential as $g_{i}(i=0,\cdots,k-1 )\rightarrow 0$ in the electric and magnetic side. Thus the gauge group $SU(N_1-1) \times \cdots \times SU(N_k-1) \times U(1)^{k-1} $ is expected to recover the dual $SU(kN_f -N_c)$ gauge group with the adjoint matter in the limit of $g_{i}(i=0,\cdots,k-1 )\rightarrow 0$. Before doing so, we will change the dynamics of the second $U(1)^{k-1}$ to the dual description with no gauge group. Here we do not expect that the recovery of the $SU(k)$ gauge group from the right-lower $U(1)^{k-1}$ in the limit of $g_{i}(i=0,\cdots,k-1 )\rightarrow 0$ modify the low-energy duality. This is only an assumption and is not justified because the non-diagonal components of the $k \times k$ adjoint fields become massless and these should be taken into account to the low-energy dynamics in the limit with $g_i \rightarrow 0$. Although this is a working assumotion, it will be found this nicely works. 
The theory can be seen as $SU(kN_f-N_c) \times U(1) \times U(1)^{k-1} \cong U(kN_f-N_c) \times U(1)^{k-1} $ gauge theory at this stage. 

With the limit of $g_i \,(i=k) \rightarrow 0$, we obtain the $U(kN_f-N_c) \times U(1)^{k-1}$ theory whose matter contents are following.

\begin{itemize}
\item $U(kN_f-N_c)$ fundamental matters: ~$N_f$ flavors $q,\tilde{q}$
\item $U(kN_f-N_c)$ adjoint matter: ~One adjoint field $Y$ which comes from the $(kN_f-N_c) \times (kN_f-N_c)$ left-upper components of the original adjoint fields $Y$. With the weak deformation $Y$ should be replaced with the vacuum expectation value.
\item The singlets $M_j,\tilde{M}_j~(j=0,\cdots,k-1)$: ~$M_j$ fields also have the $N_f \times N_f$ flavor indices. On the other hand $\tilde{M}_j$ are the flavor singlets because they are originally from the last component of $M_j$. 
\item The singlets $Y_i~(i=1,\cdots,k)$, which comes from the diagonal parts of the right-bottom $k \times k$ components of the adjoint field $Y$. These fields should be recognized as the vacuum expectation values $\braket{Y_i}$. 
\item $U(1)^{k-1}$ charged fields: $q_i , \, \tilde{q}_i~~(i=1,\cdots,k)$. We consider the gauge group $U(1)^{k-1}$ as $U(1)^k / U(1)$. In this perspective the charge assignment of $q_i,\tilde{q}_i$ are precisely the same as the 3d $\mathcal{N}=2$ mirror symmetry \cite{deBoer:1997ka,Aharony:1997bx}. They are also charged under the $U(1) \subset U(kN_f-N_c)$.
\end{itemize} 

The superpotential becomes as follows:
\begin{align}
W=\mathrm{Tr}\, Y^{k+1} +\sum_{j=0}^{k-1} M_j \tilde{q} Y^{k-1-j} q +\sum_{j=0}^{k-1}\sum_{i=1}^k \tilde{M}_j \tilde{q}_i Y^{k-1-j}_i q_i +\sum_{j=0}^{k-1} \tilde{\eta} \tilde{V}_{j,k-1-j},
\end{align}
where $Y$ in the superpotential should be constant vevs and $Y$ fields are all massive. However we loosely write the superpotential as presented above.

We define new singlet fields $S_i := \sum_{j=0}^{k-1} \tilde{M}_j Y_i^{k-1-j}~(i=1,\cdots,k)$, yielding the following superpotential.
\begin{align}
W=\mathrm{Tr} \, Y^{k+1}  +\sum_{j=0}^{k-1} M_j \tilde{q} Y^{k-1-j} q +\sum_{i=1}^k S_i \tilde{q}_i  q_i +\sum_{j=0}^{k-1} \tilde{\eta} \tilde{V}_{j,k-1-j} 
\end{align}
The $U(1)^{k-1}$ theory has the dual description by the 3d $\mathcal{N}=2$ mirror symmetry \cite{deBoer:1997ka,Aharony:1997bx} due to the appearance of the superpotential $W=\sum_{i=1}^k S_i \tilde{q}_i  q_i $. 
The mirror theory is the three-dimensional $\mathcal{N}=2$ supersymmetric $U(kN_f-N_c) \times U(1)^{\mathsf{mirror}}$ gauge theory. The field contents are summarized as follows.
\begin{itemize}
\item $U(kN_f-N_c)$ fundamentals: $N_f$ flavors $q,\tilde{q}$.
\item $U(kN_f-N_c)$ adjoint: one adjoint field $Y$
\item The $N_f \times N_f$ gauge singlets: $M_j~(j=0,\cdots,k-1)$
\item $U(1)$ $\mathcal{N}=2$ SQED with $k$ flavors $b_i, \tilde{b}_i~(i=1,\cdots,k)$
\end{itemize}
The quantum numbers of the fields are calculated by identifying the baryonic operators betwen the electric and magnetic theories. 
\begin{align}
B_{\mathsf{electric}}^{n_1,n_2,\cdots,n_k} &\equiv Q_{(0)}^{n_1} \cdots Q_{(k-1)}^{n_k}, ~~~\sum_{j=1}^k n_j =N_c \\
B_{\mathsf{magnetic}}^{m_1,\cdots,m_k} & \equiv V_{-}^{U(1)^{\mathsf{mirror}}} q_{(0)}^{m_1} \cdots q_{(k-1)}^{m_k},~~~m_j =N_f-n_{k+1-j}
\end{align}
where the magnetic baryon operators contain the Coulomb branch coordinate of the mirror $U(1)^{\mathsf{mirror}}$ because in the $U(kN_f - N_c) \times U(1)^{k-1}$ theory, $V_{-}^{U(1)^{\mathsf{mirror}}}$ is identified with $N_{-} :=q_1 q_2 \cdots q_k$ and the baryonic operators become $B_{\mathsf{magnetic}}^{m_1,\cdots,m_k}  =q_{(0)}^{m_1} \cdots q_{(k-1)}^{m_k} \prod_{i=1}^k q_i$, which are the natural baryonic operators to be identified with the electric side.

The quantum numbers are summarized as follows. Actually we have an ambiguity about the choice of the $U(1)_B$ charges to mix it with the other $U(1)$ symmetries. The $U(1)_R$ charges of the $b_i,\tilde{b}_i$ fields are different from the conventional assignment of the mirror symmetry. This is due to the presence of the superpotential $W=\sum_{j=0}^{k-1} b_{j+1} \tilde{V}^{U(kN_f-N_c)}_{k-1-j,-}$, which breaks the usual $U(1)_R$ symmetry and generate a new $U(1)_R$ symmetry at the low energy.  Notice that the $U(1)^{\mathsf{mirror}}$ symmetry is the gauging of the topological $U(1)$ symmetry corresponding to $U(1) \subset U(kN_f-N_c)$ gauge group and the $U(1) \subset U(kN_f-N_c)$ symmetry is the gauging of the topological symmetry coresponding to the $U(1)^{\mathsf{mirror}}$ gauge group. Thus we have the Chern-Simons coupling between $U(1)^{\mathsf{mirror}}$ and $ U(1) \subset U(kN_f-N_c)$.
\begin{table}[H] \caption{Quntum numbers of the mirror theory}\begin{center}
  \begin{tabular}{|c|c|c|c|c| } \hline
     & $U(kN_f-N_c) \times U(1)^{\mathsf{mirror}}$  & $U(1)_B$ & $U(1)_A$ & $U(1)_R$ \\  \hline 
    $q$  & $\left( {\tiny\yng(1)}_{\frac{1}{kN_f-N_c}},0 \right)  $  & 0 & $-1$& $-r +\frac{2}{k+1}$ \\
    $\tilde{q}$  & $ \left( \overline{{\tiny\yng(1)}}_{-\frac{1}{kN_f-N_c}},0 \right) $  & $0$& $-1$ & $-r+\frac{2}{k+1}$  \\
   $Y$ & $(\mathsf{adj.},0 )$ &  0 & 0 & $\frac{2}{k+1}$ \\  
 $b_i$  & $(1_0, 1)$ & 0 & $-N_f$ &  $-(r-1)N_f -\frac{2N_c}{k+1} +\frac{2i}{k+1}  $ \\
  $\tilde{b}_i$ & $(1_0, -1)$  & 0 & $-N_f$ & $-(r-1)N_f -\frac{2N_c}{k+1} +\frac{2i}{k+1}  $  \\
   $M_j$ & $(1_0,0) $  &  0 & 2 & $2r+\frac{2j}{k+1}$ \\ \hline
   $V_{\pm}^{U(1)^{\mathsf{mirror}}}$ & $(1_{\pm 1},0 )$  & $\mp N_c $& $kN_f$ & $(r-1)kN_f +\frac{2kN_c}{k+1}$   \\
  $\tilde{V}_{j,\pm}^{U(kN_f-N_c)}$ & $(1_0, \pm 1)$ & 0 & $N_f$ & $(r-1)N_f + \frac{2}{k+1}(N_c+1) +\frac{2j}{k+1}  $
   \\ \hline
  \end{tabular}\end{center}
\end{table}

The superpotential is 
\begin{align}
W=\mathrm{Tr} \, Y^{k+1}+ \sum_{j=0}^{k-1} M_j \tilde{q} Y^{k-1-j} q +\sum_{j=0}^{k-1} \tilde{\eta} \tilde{V}_{j,k-1-j}
\end{align}
%
The monopole operators at high energy should be identified as follows.
\begin{align}
\tilde{V}_{j,k-1-j} = b_{j+1} \tilde{V}^{U(kN_f-N_c)}_{k-1-j,-},~~(j=0,\cdots,k-1).
\end{align}
In addition to the above superpotential we have the Affleck-Harvey-Witten type superpotential \cite{Affleck:1982as} which is generated by the gauge symmetry breaking $SU(k(N_f+1) -N_c) \rightarrow U(kN_f-N_c) \times U(1)^{k-1} \underset{\mathsf{mirror}}{\approx} U(kN_f-N_c) \times U(1)^{\mathsf{mirror}} $:
\begin{align}
W_{\mathsf{AHW}} =\sum_{j=0}^{k-1} \tilde{b}_{j+1} \tilde{V}^{U(kN_f-N_c)}_{k-1-j,+}
\end{align}
Putting these into the superpotential we obtain the dual (mirror) theory,
\begin{align}
W&=\mathrm{Tr} \, Y^{k+1} + \sum_{j=0}^{k-1}M_j \tilde{q}Y^{k-1-j} q +\sum_{j=0}^{k-1} \left( b_{j+1} \tilde{V}_{k-1-j,-}^{U(kN_f -N_c)}  + \tilde{b}_{j+1} \tilde{V}_{k-1-j,+}^{U(kN_f -N_c)}   \right),
\end{align}
where $\tilde{\eta}$ is absorbed by the field rescaling.
This is precisely in accord with with the result obtained by the un-gauging technique \cite{Park:2013wta}.

\section{Summary}
In this paper we derived the duality of the three-dimensional $\mathcal{N}=2$ supersymmetric gauge theory with $N_f$ fundamental matters and one adjoint matter with a superpotential $W=\mathrm{Tr} \,X^{k+1}$ from the four-dimensional Kutasov-Schwimmer duality. We especially concentrated on $U(N_c)$, $SU(N_c)$ gauge groups and the dualities obtained here were the ones which are given by \cite{Kim:2013cma,Park:2013wta}. While in \cite{Kim:2013cma,Park:2013wta} the dualities were constructed by the generalization of the Aharony duality and by the un-gauging technique respectively, we offered the alternative derivation of these dualities in this paper. The duality for the $U(N_c)$ with an adjoint matter was derived in a similar way of the derivation of the Aharony duality from the 4d $U(N_c)$ Seiberg duality \cite{Aharony:1997gp}. Hence, the superpotential from the KK-monopoles,  Affleck-Harvey-Witten and the XYZ model from the $U(1)^2$ part play a crucial role to modify the Coulomb branch on the magnetic side. On the other hand, the $SU(N_c)$ duality with an adjoint matter is derived in a same way as the $U(N_c)$ case, where we used the $\mathcal{N}=2$ mirror symmetry which is the generalization of the duality between the SQED with $N_f=1$ and the XYZ model, which makes the Coulomb branch of the dual gauge theory lifted together with the Affleck-Harvey-Witten type superpotential.

It would be worth analyzing the derivation in this paper without the weak perturbation for the adjoint matters $X$ and $Y$, in which we will have to include the dynamics of the non-abelian gauge group $SU(k)$ for the $SU(N_c)$ duality and $U(k) \times U(k)$ for the $U(N_c)$ duality.  In deriving the duality we assumed that the $U(1)^{k-1}$ part does not enhance to $SU(k)$ in the limit of $s_i \rightarrow 0$ for $SU(N_c)$. In the $U(N_c)$ cases we relied on the similar assumption. This was only the working assumption that discarding the off-diagonal part of the adjoint fields would not be important for deriving dualities, hence it should be investigated more carefully. It would be also worth considering other gauge groups, for example $O(N_c)$, $SO(N_c)$ and $Sp(2N_c)$ with an adjoint matter.

It is important to extend the procedure of deriving the 3d dualities from 4d to the dualities with various matter fields as \cite{Csaki:2014cwa} (See \cite{Intriligator:1995ax} for examples of the 4d dualities). In \cite{Intriligator:1995ax} the symmetric tensor matter or anti-symmetric tensor matter are added to the 4d dualities and the chiral theories are also considered. It is interesting to study the corresponding 3d dualities. Although the procedure to obtain the 3d dualities is the same as ones in \cite{Aharony:2013dha,Aharony:2013kma,Csaki:2014cwa} and in this paper, it would be more and more subtle because the structure of the Coulomb branch becomes complicated. To obtain the 3d dualities, it is necessary to match the monopole operators at high- and low-energy theory when we turn on the real masses and take the low energy limit and it is very subtle task. This extension is left as a future work.

\section*{Acknowledgments}

 We would like to thank Prof. Ofer Aharony, Itamar Shamir, and Prof. Tadakatsu Sakai for helpful discussions. We thank Weizmann Institute, where the half of this work was done. This work was supported in part by Leadership Development Program for Space Exploration and Research at Nagoya University.



\begin{thebibliography}{40}
\bibitem{Seiberg:1994pq} 
  N.~Seiberg,
  ``Electric - magnetic duality in supersymmetric nonAbelian gauge theories,''
  Nucl.\ Phys.\ B {\bf 435}, 129 (1995)
  [hep-th/9411149].

\bibitem{Aharony:1997bx} 
  O.~Aharony, A.~Hanany, K.~A.~Intriligator, N.~Seiberg and M.~J.~Strassler,
  ``Aspects of N=2 supersymmetric gauge theories in three-dimensions,''
  Nucl.\ Phys.\ B {\bf 499}, 67 (1997)
  [hep-th/9703110].

\bibitem{Intriligator:1995ax} 
  K.~A.~Intriligator, R.~G.~Leigh and M.~J.~Strassler,
  ``New examples of duality in chiral and nonchiral supersymmetric gauge theories,''
  Nucl.\ Phys.\ B {\bf 456}, 567 (1995)
  [hep-th/9506148].

\bibitem{Kutasov:1995np} 
  D.~Kutasov and A.~Schwimmer,
  ``On duality in supersymmetric Yang-Mills theory,''
  Phys.\ Lett.\ B {\bf 354}, 315 (1995)
  [hep-th/9505004].

\bibitem{Kutasov:1995ve} 
  D.~Kutasov,
  ``A Comment on duality in N=1 supersymmetric nonAbelian gauge theories,''
  Phys.\ Lett.\ B {\bf 351}, 230 (1995)
  [hep-th/9503086].
  
\bibitem{Kapustin:1996nb} 
  A.~Kapustin,
  ``The Coulomb branch of N=1 supersymmetric gauge theory with adjoint and fundamental matter,''
  Phys.\ Lett.\ B {\bf 398}, 104 (1997)
  [hep-th/9611049].
 
\bibitem{Karch:1997ux} 
  A.~Karch,
  ``Seiberg duality in three-dimensions,''
  Phys.\ Lett.\ B {\bf 405}, 79 (1997)
  [hep-th/9703172].
  
\bibitem{Aharony:1997gp} 
  O.~Aharony,
  ``IR duality in d = 3 N=2 supersymmetric USp(2N(c)) and U(N(c)) gauge theories,''
  Phys.\ Lett.\ B {\bf 404}, 71 (1997)
  [hep-th/9703215].
  
\bibitem{Giveon:2008zn} 
  A.~Giveon and D.~Kutasov,
  ``Seiberg Duality in Chern-Simons Theory,''
  Nucl.\ Phys.\ B {\bf 812}, 1 (2009)
  [arXiv:0808.0360 [hep-th]].
  
\bibitem{Niarchos:2008jb} 
  V.~Niarchos,
  ``Seiberg Duality in Chern-Simons Theories with Fundamental and Adjoint Matter,''
  JHEP {\bf 0811}, 001 (2008)
  [arXiv:0808.2771 [hep-th]].
  
\bibitem{Niarchos:2009aa} 
  V.~Niarchos,
  ``R-charges, Chiral Rings and RG Flows in Supersymmetric Chern-Simons-Matter Theories,''
  JHEP {\bf 0905}, 054 (2009)
  [arXiv:0903.0435 [hep-th]].

\bibitem{Intriligator:2013lca} 
  K.~Intriligator and N.~Seiberg,
  ``Aspects of 3d N=2 Chern-Simons-Matter Theories,''
  JHEP {\bf 1307}, 079 (2013)
  [arXiv:1305.1633 [hep-th]].
  
\bibitem{Aharony:2013dha} 
  O.~Aharony, S.~S.~Razamat, N.~Seiberg and B.~Willett,
  ``3d dualities from 4d dualities,''
  JHEP {\bf 1307}, 149 (2013)
  [arXiv:1305.3924 [hep-th]].
  
\bibitem{Aharony:2013kma} 
  O.~Aharony, S.~S.~Razamat, N.~Seiberg and B.~Willett,
  ``3$d$ dualities from 4$d$ dualities for orthogonal groups,''
  JHEP {\bf 1308}, 099 (2013)
  [arXiv:1307.0511 [hep-th]].
  
\bibitem{Kim:2013cma} 
  H.~Kim and J.~Park,
  ``Aharony Dualities for 3d Theories with Adjoint Matter,''
  JHEP {\bf 1306}, 106 (2013)
  [arXiv:1302.3645 [hep-th]].
  
\bibitem{Park:2013wta} 
  J.~Park and K.~J.~Park,
  ``Seiberg-like Dualities for 3d N=2 Theories with SU(N) gauge group,''
  arXiv:1305.6280 [hep-th].
  
\bibitem{deBoer:1997ka} 
  J.~de Boer, K.~Hori, Y.~Oz and Z.~Yin,
  ``Branes and mirror symmetry in N=2 supersymmetric gauge theories in three-dimensions,''
  Nucl.\ Phys.\ B {\bf 502}, 107 (1997)
  [hep-th/9702154].

\bibitem{Intriligator:1996ex} 
  K.~A.~Intriligator and N.~Seiberg,
  ``Mirror symmetry in three-dimensional gauge theories,''
  Phys.\ Lett.\ B {\bf 387}, 513 (1996)
  [hep-th/9607207].
  
\bibitem{Callias:1977kg} 
  C.~Callias,
  ``Index Theorems on Open Spaces,''
  Commun.\ Math.\ Phys.\  {\bf 62}, 213 (1978).

\bibitem{Weinberg:1979zt} 
  E.~J.~Weinberg,
  ``Fundamental Monopoles and Multi-Monopole Solutions for Arbitrary Simple Gauge Groups,''
  Nucl.\ Phys.\ B {\bf 167}, 500 (1980).
  
\bibitem{Weinberg:1982ev} 
  E.~J.~Weinberg,
  ``Fundamental Monopoles in Theories With Arbitrary Symmetry Breaking,''
  Nucl.\ Phys.\ B {\bf 203}, 445 (1982).
  
\bibitem{Atiyah:1963zz} 
  M.~F.~Atiyah and I.~M.~Singer,
  ``The index of elliptic operators on compact manifolds,''
  Bull.\ Am.\ Math.\ Soc.\  {\bf 69}, 422 (1969).

\bibitem{Atiyah:1968mp} 
  M.~F.~Atiyah and I.~M.~Singer,
  ``The Index of elliptic operators. 1,2,3''
  Annals Math.\  {\bf 87}, 484 (1968).
  
\bibitem{Poppitz:2008hr} 
  E.~Poppitz and M.~Unsal,
  ``Index theorem for topological excitations on R**3 x S**1 and Chern-Simons theory,''
  JHEP {\bf 0903}, 027 (2009)
  [arXiv:0812.2085 [hep-th]].
  

\bibitem{deBoer:1997kr} 
  J.~de Boer, K.~Hori and Y.~Oz,
  ``Dynamics of N=2 supersymmetric gauge theories in three-dimensions,''
  Nucl.\ Phys.\ B {\bf 500}, 163 (1997)
  [hep-th/9703100].

\bibitem{Affleck:1982as} 
  I.~Affleck, J.~A.~Harvey and E.~Witten,
  ``Instantons and (Super)Symmetry Breaking in (2+1)-Dimensions,''
  Nucl.\ Phys.\ B {\bf 206}, 413 (1982).
  
\bibitem{Csaki:2014cwa} 
  C.~Cs$\mathrm{\acute{a}}$ki, M.~Martone, Y.~Shirman, P.~Tanedo and J.~Terning,
  ``Dynamics of 3D SUSY Gauge Theories with Antisymmetric Matter,''
  arXiv:1406.6684 [hep-th].
  

\end{thebibliography}
\end{document}